\documentclass[RNAAS]{aastex62}

\shortauthors{Plewa et al.}

\begin{document}

\title{Optical distortion in the NACO imager}

\correspondingauthor{P. M. Plewa}
\email{pmplewa@mpe.mpg.de}

\author{P. M. Plewa}
\affil{Max-Planck-Institut f\"ur Extraterrestrische Physik, Garching, Germany}

\author{S. Gillessen}
\affil{Max-Planck-Institut f\"ur Extraterrestrische Physik, Garching, Germany}

\author{M. Baub\"ock}
\affil{Max-Planck-Institut f\"ur Extraterrestrische Physik, Garching, Germany}

\author{J. Dexter}
\affil{Max-Planck-Institut f\"ur Extraterrestrische Physik, Garching, Germany}

\author{F. Eisenhauer}
\affil{Max-Planck-Institut f\"ur Extraterrestrische Physik, Garching, Germany}

\author{S. von Fellenberg}
\affil{Max-Planck-Institut f\"ur Extraterrestrische Physik, Garching, Germany}

\author{F. Gao}
\affil{Max-Planck-Institut f\"ur Extraterrestrische Physik, Garching, Germany}

\author{R. Genzel}
\affil{Max-Planck-Institut f\"ur Extraterrestrische Physik, Garching, Germany}

\author{M. Habibi}
\affil{Max-Planck-Institut f\"ur Extraterrestrische Physik, Garching, Germany}

\author{A. Jimenez-Rosales}
\affil{Max-Planck-Institut f\"ur Extraterrestrische Physik, Garching, Germany}

\author{T. Ott}
\affil{Max-Planck-Institut f\"ur Extraterrestrische Physik, Garching, Germany}

\author{O. Pfuhl}
\affil{Max-Planck-Institut f\"ur Extraterrestrische Physik, Garching, Germany}

\author{I. Waisberg}
\affil{Max-Planck-Institut f\"ur Extraterrestrische Physik, Garching, Germany}

\author{F. Widmann}
\affil{Max-Planck-Institut f\"ur Extraterrestrische Physik, Garching, Germany}

\section{A distortion correction for the S13 camera}

\begin{figure}[ht!]
\centering
\includegraphics[width=0.8\linewidth]{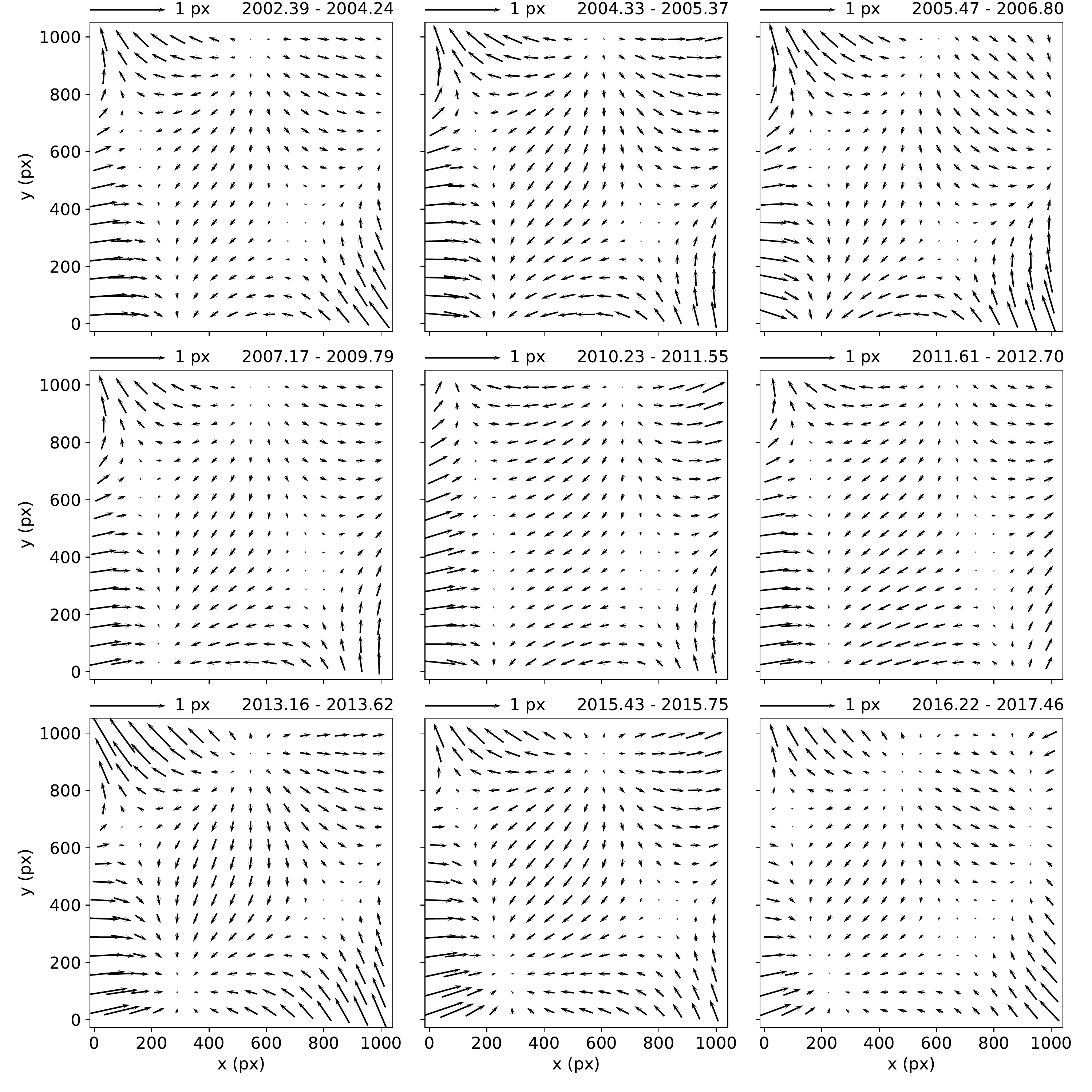}
\caption{Distortion corrections for images taken with the S13 camera of NACO, for different periods between 2002 and 2017.}
\label{figure}
\end{figure}

We present a set of distortion solutions (Fig.~\ref{figure}) that may be used to correct geometric optical distortion in images taken with the S13 (narrow-field) camera of the NACO adaptive optics imager \citep{1998SPIE.3353..508R,1998SPIE.3354..606L}, complementing the distortion corrections we have previously derived for the S27 (wide-field) camera \citep{2015MNRAS.453.3234P}. As a result of the S13 camera distortions, observed positions on the detector are offset from the true positions by about 1.2-2.3~mas on average, and as much as 7~mas in the corners of the detector, at a pixel scale of $\sim$13~mas. The overall distortion pattern is complex but generally static, yet appears to have changed abruptly at least 9 times over the time that NACO has been in regular operation at the VLT, between 2002 and 2017.

\section{Mapping the image distortion}

We derive the new distortion corrections from near-infrared, high-angular resolution imaging observations of the Galactic Center, which continue to be routinely performed with NACO in the H- and Ks-bands, using both the S27 and the S13 camera, with the primary goal of monitoring stellar motions around the central supermassive black hole \citep[e.g.][]{2009ApJ...692.1075G,2017ApJ...837...30G}. The existing set of distortion corrections for the S27 camera has allowed us to create a catalog of about 2000 astrometric reference stars in the inner nuclear star cluster, for which we can measure precise proper motions in an approximately distortion-free reference frame \citep[see][]{2015MNRAS.453.3234P}. For every suitable image taken with the S13 camera, we determine the optimal affine transformation (i.e. translation, rotation, scaling and shear operations; accounting for differential refraction, as well as other linear effects) to align the astrometric positions of these reference stars, at the respective times of observations, with their pixel positions as measured on the detector. The map of transformation residuals, i.e. the remaining difference between aligned positions, is a vector field that represents the higher-order distortions still present in each image. Due to the high stellar density in the Galactic Center, we can match as many as a few hundred stars per image, to create a combined distortion map for each observing night that is sampled at several thousand positions across the detector. A self-calibration approach \citep[e.g. as used by][]{2003PASP..115..113A} is not feasible, despite the high stellar density, because the requirement that many of the same stars are placed (at various orientations) in many different regions on the detector is almost never met, due to the specific dithering scheme employed for the past Galactic Center monitoring observations.

\section{The distortion model}

The distortion maps, after smoothing, are fit by a model that fully captures the spatial variability of the image distortion in 20 free parameters, and describes the distortion maps in terms of a basis set of orthonormal vector polynomials \citep{2007OExpr..1518014Z,2008OExpr..16.6586Z}. In addition to enabling the explicit correction of optical distortion in other NACO images, the fit results justify using cubic transformations to place stellar positions measured on widely dithered images of the Galactic Center into a common (relative) astrometric reference frame, which has been the standard practice to account for unknown image distortions in several previous studies \citep[e.g.][]{2009ApJ...692.1075G,2017ApJ...837...30G,2010MNRAS.401.1177F}. By a visual inspection assisted by clustering algorithms, we can identify several abrupt changes in the image distortion that lead us to create 9 distinct, averaged distortion solutions, each of which is applicable for periods of time ranging from 4 to 31~months, over which the image distortion appears to be static, to within 10\%. There are no indications for additional systematic changes in these solutions, for instance due to the different filters used. All solutions can be found online, at \url{http://www.mpe.mpg.de/ir/gc/distortion}, in the form of FITS files.

\end{document}